\documentclass[12pt]{article}
\def\today{}
\usepackage{graphicx}
\usepackage{dcolumn}
\usepackage{bm}
\usepackage{latexsym}

\begin{document}
\title{Time-dependent electron transport \\ through the multi-terminal quantum dot}
\author{R. Taranko\footnote{corresponding author, e-mail:
taranko@tytan.umcs.lublin.pl} \, T. Kwapi\'nski and E. Taranko \\ \\
Institute of Physics, M. Curie-Sk\l odowska University \\
Pl. M. Curie-Sk\l odowskiej 1, 20-031 Lublin, Poland}

\maketitle

\date{\today}

\vspace{0.5cm}

\begin{abstract}
We consider the time-dependent electron transport through a quantum dot connected to multiple leads in the
presence of the additional over-dot (bridge) tunnelling channels by using the evolution operator technique. Each terminal and
quantum dot are disturbed by an external oscillating field resulting in a time-dependence of the corresponding energy levels.
The final analytical expressions for the currents flowing in the system are given assuming the wide-band limit approximation. We
investigate also the transient-current characteristics in the case of the rectangular-pulse modulations imposed on the dot-lead
barriers. The time-averaged current and its derivative with respect to the gate voltage have been calculated for a wide range of
parameters.
\end{abstract}

\section{\label{sec1} Introduction }

The advance of experimental techniques on a nanometer scale has enhanced the interest on the electronic transport through
quantum dots. Especially interesting are the transport properties of a quantum dot (QD) under the influence of external
time-dependent fields. The microwave fields applied to different parts of the system under consideration modify the QD charge
and tunneling current. New effects have been observed and theoretically described, e.g. photon-assisted tunneling through small
QD-s with well-resolved discrete energy states, photon-electron pumps  and others \cite{pla}. One can investigate the current
flowing through a QD under periodic (non-periodic) modulation of the tunneling barriers or under harmonic modulation of the
electron energy levels in both (say, left and right) electron reservoirs, e.g.  \cite{zha, jau}. One of the important problem of
mesoscopic physics is the interference of the charge carriers. This interference appears when two (or more) transmission
channels for tunneling electron exist. The experimental situation in which the destructive interference may occur can be
realized in the scanning tunneling microscope (STM) or in multi-terminal QD system.

In this paper we generalize models existing in the literature and consider the QD connected with three leads with additional
over-dot (bridge) tunneling channel between leads, e.g. \cite{ma, kwa, kwa2,kwa3}. This case corresponds to the possible STM
experimental setup in which the QD placed between two leads - say the right leads, can be probed by means of the additional
electrode (tip) - say the left lead. In such configuration the additional channels for the electron transfer between STM tip and
the right and left leads exist.  We consider the system driven out of the equilibrium by means of a dc voltage bias and
time-dependent external fields. To treat this nonequilibrium, time-dependent electron transport process we use the evolution
operator technique and find the final expression for the current flowing in the system in terms of the appropriate matrix
elements of this operator.

\section{\label{sec1}Model and calculation method }

The Hamiltonian of the QD coupled through the tunneling barriers $V_{\vec k_id}$ ($i=1,2,\ldots,N$) to $N$ metal leads with
chemical potential $\mu_i$ can be written as $H=H_0(t)+V(t)$, where
\begin{equation}
H_0(t) = \sum^N_{i=1} \sum_{\vec k_i} \varepsilon_{\vec k_i}(t) c^+_{\vec k_i} c_{\vec k_i} + \varepsilon_0(t) c^+_d c_d \,,
\label{eq1}
\end{equation}
\begin{equation}
V(t) =\sum^N_{i,j=1 \,;\, i<j} \,\,\sum_{\vec k_i, \vec k_j} V_{\vec k_i \vec k_j}(t) c^+_{\vec k_i}c_{\vec k_j} + {\rm h.c.} +
\sum^N_{i=1} \sum_{\vec k_i} V_{\vec k_i d}(t) c^+_{\vec k_i} c_d + {\rm h.c.} \label{eq2}
\end{equation}
within usually used notation. For simplicity the dot is characterized only by the single level $\varepsilon_0$ and we have
neglected the inta-dot electron-electron interaction. We assume the microwave fields applied to the leads and the QD as follows:
$\varepsilon_{\vec k_i}(t) = \varepsilon_{\vec k_i} + \Delta_i\cos\omega t$ and $\varepsilon_0(t) = \varepsilon_0 + \Delta_0
\cos\omega t$. The dynamical evolution of the charge localized on the QD and the current flowing in the system can be described
in terms of the time-evolution operator $U(t,t_0)$, \cite{kwa, kwa2, kwa3} and, for example, the tunneling current flowing from
the $i$-th lead can be obtained from the time derivative of the total number of electrons in this lead, $j_i(t) = -edn_i(t)/dt$
(cf. \cite{jau}), where
\begin{eqnarray}
{n_i(t)}=\sum_{\vec k_i} n_{\vec k_i}(t) = \sum_{\vec k_i}
 [n_d(t_0)|U_{\vec k_i d}(t,t_0)|^2+
\sum^n_{j=1} \sum_{\vec k_j} n_{\vec k_j}(t_0)|U_{\vec k_i,\vec k_j}(t,t_0)|^2]\,.\, \label{eq7}
\end{eqnarray}
Here $U_{\vec k_i d}(t,t_0)$ and $U_{\vec k_i \vec k_j}(t,t_0)$ denote the matrix elements of $U(t,t_0)$ calculated within the
basis functions containing the single-particle functions $|d\rangle$ and $|\vec k_i\rangle$ corresponding to the QD and $i$-th
metal lead, respectively. $n_d(t_0)$ and $n_{\vec k_i}(t_0)$ represent the initial filling of the corresponding single-particle
states.

Assuming the wide-band limit and after lengthy calculations one obtains the average current leaving e.g. the left lead (QD
coupled with three leads $L,R_1$ and $R_2$) in the form:

\begin{eqnarray}
\langle j_L\rangle &=& {2x^2 \over (1+2x^2)^2} (2\mu_L-\mu_{R_1}-\mu_{R_2})+ \nonumber\\ && {\Gamma \over \pi(1+2 x^2)^3}
\left[-{1 \over 3} (1-13 x^2+4x^4){\rm{Im}} \Phi+2x(1-2 x^2) \rm{Re}\Phi \right] ,
\end{eqnarray}
where
\begin{eqnarray}
\Phi=2 \int f_L(\varepsilon) \langle A_L(\varepsilon)\rangle d\varepsilon - \sum_{i=R_1,R_2} \int f_i(\varepsilon) \langle
A_i(\varepsilon)\rangle d\varepsilon ,
\end{eqnarray}
and
\begin{eqnarray}
\langle A_i(\varepsilon)\rangle = \sum_k J_k^2 \left( {\Delta_0-\Delta_i \over \omega} \right) \left(
\varepsilon-\varepsilon_0-\omega k+{2 \Gamma x \over 1+2x^2}+i {3\Gamma \over 2(1+2x^2)} \right)^{-1} .
\end{eqnarray}
Here $x=\pi V_{LR}/D$, $\Gamma$ is the coupling strength between QD and leads (see the next section) and $J_i(x)$ denotes the
Bessel function.

\section{\label{sec3} Results and discussion}
We consider the QD coupled with two and three metal leads with the additional over-dot (bridge) couplings between leads. The
time-dependent currents are calculated in the case when the periodic rectangular-pulse external field is applied to each QD-lead
barrier. In such a case we integrate numerically the corresponding set of the differential equations for the matrix elements of
the evolution operator. We consider also the time-averaged values for
 currents and conductance in the case of harmonic modulation of
 the system parameters. We assume the temperature $T = 0$~K and we take for $V_{\vec k_i, \vec k_j}$ the
values comparable with $V_{\vec k_\alpha d}$. We estimated $V_{\vec k_\alpha d}$ (assuming its $\vec k$-independence, $V_{\vec
k_\alpha d} \equiv V_\alpha = V$) using the relation $\Gamma_\alpha = 2\pi|V_\alpha|^2/D_\alpha$, where $D_\alpha$ is the
$\alpha$-lead's bandwidth and $D_\alpha = 100~\Gamma_\alpha$ ($\Gamma_L = \Gamma_R = \Gamma$, $D_L = D_R = D$ was assumed). In
our calculations we assumed $e=1$, all energies are given in $\Gamma$ units, time in $\hbar /\Gamma$ units, the current, its
derivatives and frequency are given in $e\Gamma/\hbar, e^2\Gamma/\hbar$ and $\Gamma/\hbar$ units, respectively.

\begin{figure}[h]
\begin{center}
 \resizebox{0.6\columnwidth}{!}{
  \includegraphics{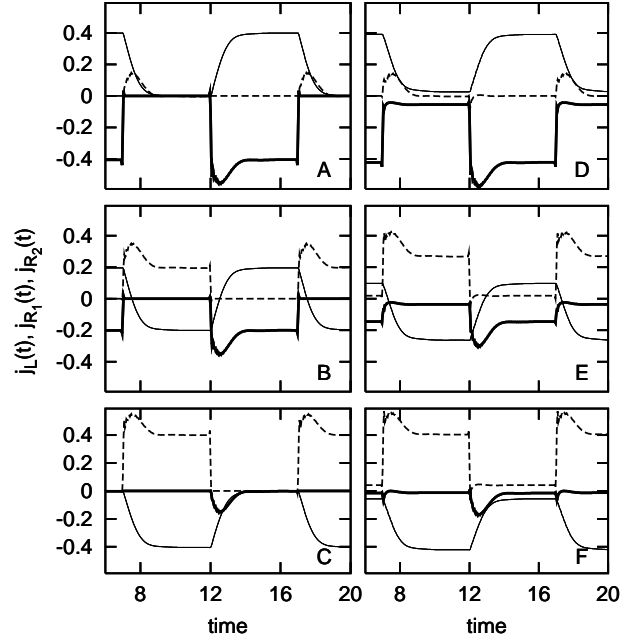}}
\end{center}
 \caption{\label{Fig1} The time-dependent current flowing in the system of a QD
coupled with three leads: $L$, $R_1$ and $R_2$. The $L$-lead is coupled with the QD only - the left panels and with the QD and
two other leads, $V_{LR_1}=V_{LR_2}=4$ - the right panels. The couplings between the QD and $R_1, R_2$ leads is changed
periodically. The upper, middle and lower panels correspond to $\mu_L=3,0$ and -3, respectively. $\mu_{R_1}=-\mu_{R_2}=3,
\varepsilon_d=0$. The thin, thick and broken curves correspond to $j_L, j_{R_1}$ and $j_{R_2}$ currents, respectively.}
\end{figure}
In Fig. \ref{Fig1} we show the currents for the QD coupled with three leads $R_1, R_2$ and $L$ with the additional couplings
$V_{LR_1}$ and  $V_{LR_2}$. The barriers QD-$R_1$ lead and QD-$R_2$ lead are changing in time according to a periodic
rectangular-pulse external field with the period $T=10$, which is applied to each barrier and these two fields are out of phase
with a phase difference of $\pi$. The third lead ($L$) is connected with the QD through the time-independent barrier. We checked
that the QD charge hardly depends on the additional $V_{LR}$ couplings. Although the QD charge is almost insensitive to the
additional over-dot couplings the currents demonstrate such dependence. Especially visible are the differences for the case when
the chemical potential $\mu_L$ of the third electrode $L$ lies between chemical potentials of two other leads, see Fig.
\ref{Fig1} B,E. For other values of $\mu_L$ (relative to $\mu_{R_1}$ and $\mu_{R_2}$) the influence of the over-dot tunneling
channels considered here is smaller. Note, that after the abrupt changing of the coupling strength the currents $j_L, j_{R_1}$
and $j_{R_2}$ are rapidly changed too, and after a short time reach the steady values. The QD coupled to three leads could be
considered as the three-state system. As we see in Fig. \ref{Fig1} by changing the coupling strength the current changes
its value from e.g. zero to the positive (negative) value or from the negative to the positive value and vice versa, depending
on the chemical potentials of all leads. The additional couplings
 between leads modify the currents flowing in this system.

\begin{figure}[h]
\begin{center}
 \resizebox{0.74\columnwidth}{!}{
  \includegraphics{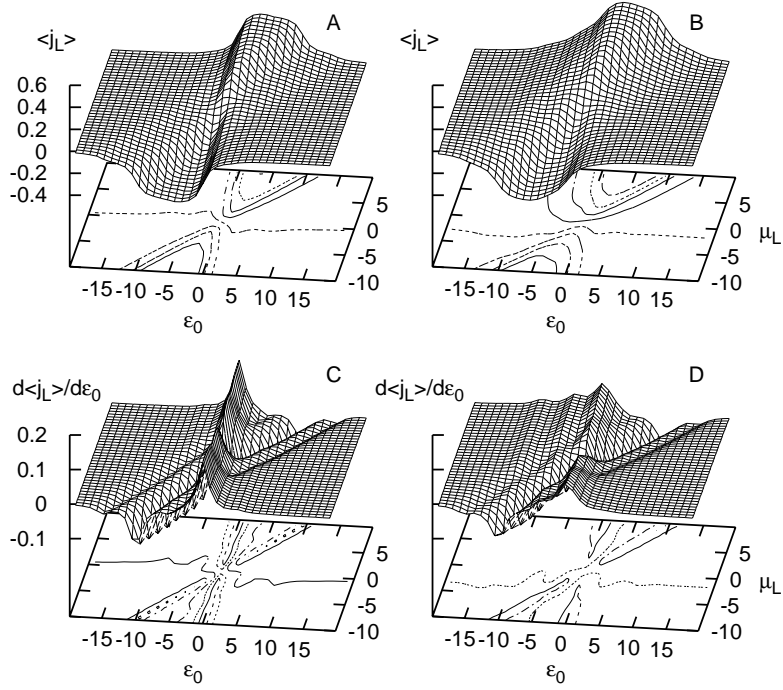}}
\end{center}
 \caption{\label{Fig2} The time-averaged current $\langle j_L \rangle$ flowing from the left lead (upper panels) and its derivative with respect to $\varepsilon_0$
 (lower panels) as a function of $\varepsilon_0$ and $\mu_L$. The left panels correspond to the QD coupled with two leads:
 $\Delta_L=8, \Delta_0=4, \Delta_R=2$, $\mu_R=0$ and the right panels correspond to three-terminal system:  $\Delta_L=8, \Delta_0=4, \Delta_{R_1}=2$,
$\Delta_{R_2}=-2$, $\mu_{R_1}=0$, $\mu_{R_2}=-4$ and  $\omega=5, V_{LR_1}=V_{LR_2}=0$, $\Gamma=1$. }
\end{figure}
In the next step we show results for time-averaged current and its derivative with respect to $\varepsilon_0$, calculated as the
functions of $\varepsilon_0$ and $\mu_L$, for the case of the QD coupled with two leads, Fig. \ref{Fig2}A,C and coupled with
three leads, Fig. \ref{Fig2}B,D. In principle, inclusion of the third electrode does not introduce significant changes to the current
$\langle j_L \rangle$, and only the dependence on the gate voltage is more "diffusive". More transparent changes are visible on
$d\langle j_L \rangle/d\varepsilon_0$ curves, calculated as  functions of $\varepsilon_0$ and $\mu_L$ (Fig. \ref{Fig2}C,D). Now,
for the three-terminal QD system we observe three distinct enhancements regions going along $\varepsilon_0$-axis at constant
$\mu_L$ (see, for example, the changes at $\mu_L=10$) whereas for the QD coupled with two electrodes we observe only one
corresponding peak.

\begin{figure}[h]
\begin{center}
 \resizebox{0.75\columnwidth}{!}{
  \includegraphics{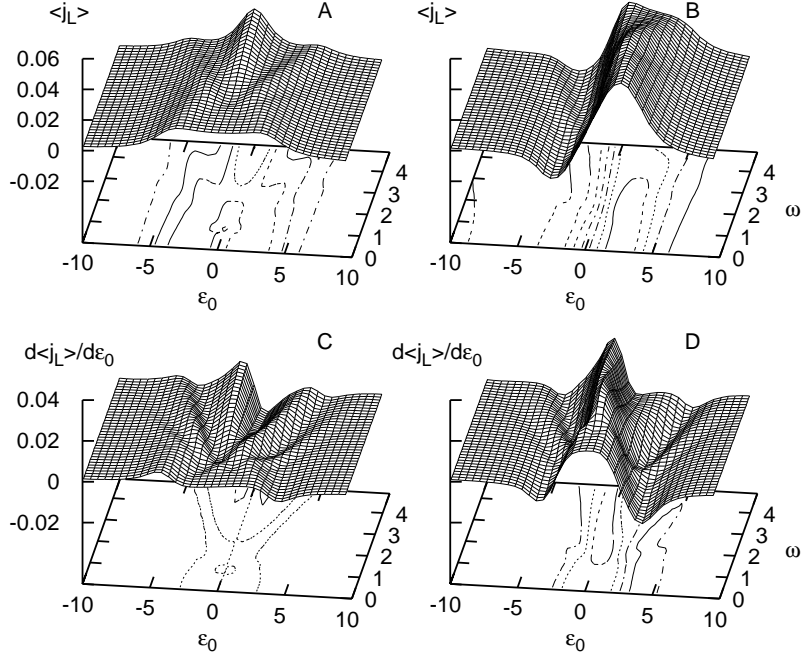}}
\end{center}
 \caption{\label{Fig3} The time-averaged current $<j_L>$ (upper panels) and its derivative with respect to $\varepsilon_0$
 (lower panels) as a function of $\varepsilon_0$ and the frequency $\omega$. $\Delta_L=8, \Delta_0=4, \Delta_{R_1}=0$,
  $\mu_L=0.2, \mu_{R_1}=-0.2$, and $\mu_{R_2}=0$. The amplitudes of the oscillation of the third electrode are $\Delta_{R_2}=0$
   (left panels) and $\Delta_{R_2}=2$ (right panels). $V_{LR_1}=V_{LR_2}=0$, $\Gamma=1$. }
\end{figure}

More dramatic differences can be observed in the three-terminal QD system when we introduce changes for parameters of the one
electrode, only. In Fig. \ref{Fig3} we show $\langle j_L \rangle$ and $d \langle j_L \rangle/d\varepsilon_0$ for two different
sets of parameters. The panels A and C correspond to the system in which one of the right lead is not affected by the
time-dependent field, $\Delta_{R_2}=0$, whereas the panels B and D show results for the case $\Delta_{R_2}=2$. The results
presented on the panels A and C are, in fact, very similar to the results obtained for two-terminal QD with $\Delta_L=8$,
$\Delta_0=4$ and $\Delta_{R_1}=0$ (not shown here). So, depending on the amplitude of oscillations of the microwave field
applied to the additional third electrode, the resulting current flowing out of the left electrode can differ in magnitude and
functional dependence on $\varepsilon_0$ and $\omega$ from that ones for slightly different parameters of the third lead.

In summary, we considered the currents flowing in the system of the multiterminal QD using the evolution operator approach. The
influence of the external harmonic microwave fields and rectangular-pulse modulations of the dot-lead barriers on the currents
was investigated.

{\bf Acknowledgements:} The work by one of us (RT) has been partially supported by the KBN grant No. PBZ-MIN-008/P03/2003.



\end{document}